\documentclass[12pt]{article}
\usepackage{a4wide}
\usepackage{amssymb, amsmath,cite}
\usepackage[normalem]{ulem}
\usepackage[utf8]{inputenc}
\usepackage[colorlinks]{hyperref}
\usepackage[usenames,dvipsnames]{color}
\hypersetup{
     breaklinks=true,
    pdfstartview={FitH},    
    colorlinks=true,       
    linkcolor=blue,          
    citecolor=red,        
    filecolor=magenta,      
    urlcolor=blue,           
    anchorcolor=green,      
    linktocpage=true
}

\begin{document}
{\renewcommand{\thefootnote}{\fnsymbol{footnote}}
		
\begin{center}
{\LARGE The trans-Planckian censorship conjecture from\\ the swampland distance conjecture} 
\vspace{1.5em}

Suddhasattwa Brahma\footnote{e-mail address: {\tt suddhasattwa.brahma@gmail.com}}
\\
\vspace{0.5em}
 Department of Physics, McGill University\\
 Montr\'eal, QC H3A 2T8, Canada\\
\vspace{1.5em}
\end{center}
}
	
\setcounter{footnote}{0}

\newcommand{\bea}{\begin{eqnarray}}
\newcommand{\eea}{\end{eqnarray}}
\renewcommand{\d}{{\mathrm{d}}}
\renewcommand{\[}{\left[}
\renewcommand{\]}{\right]}
\renewcommand{\(}{\left(}
\renewcommand{\)}{\right)}
\newcommand{\nn}{\nonumber}
\newcommand{\Mpl}{M_{\textrm{Pl}}}
\def\H{H_f}
\def\V{\mathrm{V}}
\def\e{\mathrm{e}}
\def\be{\begin{equation}}
\def\ee{\end{equation}}

\begin{abstract}
\noindent Starting from the swampland distance conjecture, and using the species bound for a large number of weakly-coupled particles, we give a derivation of the recently proposed trans-Planckian censorship conjecture. Our argument demonstrates how a quantum gravity principle \textit{requires} that trans-Planckian quantum fluctuations should never cross the Hubble horizon. We also comment on how logarithmic corrections to the de-Sitter conjecture arise naturally from such an approach when one relaxes the requirement of traversing parametrically large distances on field space.
\end{abstract}

\section{Introduction}

Recently, it has been argued that in any consistent effective field theory (EFT), there should not be such an amount of cosmic expansion that a trans-Planckian mode crosses the Hubble horizon and, consequently, becomes classical \cite{TCC1}. In other words, the trans-Planckian censorship conjecture (TCC) states that, for an EFT to \textit{not} be in the swampland, trans-Planckian quantum fluctuations should remain quantum and never cross the Hubble horizon \cite{TCC1}. This elevated the well-known `trans-Plackian problem' of inflationary cosmology \cite{TP1,TP2, TP3, TP4, TP5, TP6, TP7} to the level of a so-called swampland conjecture --  the general requirements for an EFT to have a consistent UV completion \cite{Swampland1, Swampland2, Swampland3}.

In inflation \cite{Inf1, Inf2, Inf3, Inf4, Inf5, Inf6}, one traces back macroscopic classical inhomogeneities as originating from quantum vacuum fluctuations (see, for instance \cite{Review1}) by trusting the EFT on quasi-de Sitter (dS) spacetimes to early times. However, if inflation lasts for a long time, then it is entirely possible that even a trans-Planckian quantum fluctuation would eventually cross the Hubble horizon and become classical. Since the objective of the TCC is to prevent this from happening, it puts an upper bound on the duration of the inflatonary era, such that \cite{TCC1}
\begin{eqnarray}\label{TCCorg}
e^N < \Mpl/\H \;\;\Rightarrow\; \int_{t_i}^{t_f} H \d t < \ln\(\frac{\Mpl}{\H}\)\,,
\end{eqnarray}
where $N$ is the number of $e$-folds of inflation and $\H$ denotes the Hubble parameter at the end of inflation. Assuming the TCC to be true, one can easily get an upper bound on the energy scale of inflation $\rho_\text{inf}^{1/4} < 3\times 10^{-10}\Mpl$ and, as a result, on the tensor-to-scalar ratio $r < 10^{-30}$ \cite{TCC2}. As also emphasized in \cite{TCC2}, these bounds hold quite generally for all models of inflation unless one assumes a different mechanism for the production of primordial gravitational waves \cite{Suddho} or introduces drastic changes to the cosmic history \cite{TCC_Cosmo1}\footnote{For other cosmological implications of the TCC, see \cite{TCC_Cosmo2, TCC_Cosmo3, TCC_Cosmo4, TCC_Cosmo5, TCC_Cosmo6, TCC_Cosmo7, TCC_Cosmo8, TCC_Cosmo9, TCC_Cosmo10, TCC_Cosmo11,Geng:2019phi}}. In other words, unless one introduces such radical modifications, observation of primordial tensor modes would rule out inflation, assuming the TCC to be correct. Moreover, the TCC also leads to an extreme fine-tuning problem for single-field models of inflation \cite{TCC2}. 

In light of the fact that the TCC puts into question one of our most successful paradigms of early universe cosmology -- namely, inflation -- and forces us to look at other alternatives \cite{Alternative}, it is certainly pertinent to ask why should one trust the TCC? The fundamental argument for it, as presented in \cite{TCC1}, is that if trans-Planckian modes were to cross the Hubble horizon and freeze out, then we would have observation of classical inhomogeneities which originate from trans-Planckian quantum fluctuations. This would imply that one can push inflation, as an EFT, beyond Planck scales which is the same as assuming that spacetime would remain a smooth continuum beyond such scales. Since it is well-known that imposing a naive UV cut-off to integrate out the trans-Planckian modes, in order to systematically derive a low-energy EFT, does not work for expanding backgrounds \cite{Weiss}, thus it makes sense to require that quantum fluctuations on scales smaller than the Planck length should, at the very least, be `hidden' by the Hubble horizon in analogy with Penrose's `Cosmic Censorship Hypothesis' \cite{Penrose1, Penrose2}. 

In this work, we shall show that one can arrive at the TCC (as appropriate for slow-roll scalar potentials) independent of the above motivations from cosmology. The swampland program has already established that many of these conjectures which seem to be independent and unrelated to each other on first sight, appear to be part of a deeper, inter-connected structure on closer examination \cite{SWC1, SWC2, SWC3, SWC4, SWC5, SWC6,SWC7, SWC8, SWC9, SWC10}\footnote{These are a few examples where such connections have been revealed; see, for instance, \cite{Swampland2} and references therein for a more complete list.}. A nice example of this would be \cite{SWC1}, in which it was shown that the (refined) dS conjecture follows from the swampland distance conjecture (SDC) \cite{Swampland1,SWC2,SWC3,SDC1,SDC2,SDC3} in parametrically controlled regimes of string theory, assuming Bousso's covariant entropy bound \cite{CEB1,CEB2} to hold. Although the dS conjecture \cite{SWC1,dSC1, dSC2, dSC3, dSC4, dSCnew, Dvali2, Dvali1} was originally motivated by the difficulty of constructing dS vacua in string theory, it was much more satisfying to get an argument for it in terms of a more fundamental quantum gravity consideration, namely, the entropy bound for quasi-dS spacetimes in this case. (Other arguments for the dS conjecture to arise from more fundamental aspects of quantum gravity was given in \cite{Dvali1, Dvali2}.)

In a similar vein, we shall show that the TCC can be derived starting from a well-explored aspect of quantum gravity -- the SDC -- assuming some scaling for the cut-off according to the species bound \cite{SPB1, SPB2, SPB3, SPB4}. Since the SDC has been tested extensively in concrete examples coming from string theory (\textit{e.g.,} \cite{SDC1,SDC2,SDCTest1, SDCTest2, SDCTest3,SDCTest4, SDCTest5}), our derivation makes the TCC much more robust, and consequently, its implications for cosmology more unavoidable. As an aside, it shall also demonstrate how the logarithmic corrections \cite{TCC1, dSlog} naturally arise in the dS conjecture, starting from the SDC and using the species bound if one does not require that the field excursion $|\Delta\varphi| \rightarrow \infty$. To the best of our knowledge, this point has been overlooked in previous derivations of the dS conjecture from similar arguments such as in \cite{dSfromSDC}.

Since it is rather strange that an abstract quantum gravity principle can give rise to the TCC, which is often formulated as a statement about quantum fields on quasi-dS spacetime, let us first give a brief overview of our argument. As expressed in \eqref{TCCorg}, the TCC is a statement about the maximum number of $e$-folds one can have during an accelerated phase of expansion. As mentioned above, this upper bound on the duration was motivated in \cite{TCC1, TCC2} by requiring that trans-Planckian quantum modes never cross the Hubble horizon. Starting from the SDC, and using the species bound, one finds an upper bound on the UV cut-off of the weakly-coupled gravitational theory. This cut-off decreases exponentially on field space if there is a field-dependent tower of states descending from the UV (as is the case for the SDC). Consequently, one gets an upper bound for the duration of an accelerated phase by imposing that the Hubble parameter, at the end of this phase, is below the cut-off of the theory. This, in short, forms the crux of our argument and shows how this requires that trans-Planckian quantum fluctuations should never cross the Hubble horizon based on a completely different, and well-tested, string theory argument. As shall be emphasized later, the crucial thing for our reasoning to go through is that the exponent $\alpha$ from the SDC and the one which appears for the TCC are both $\mathcal{O}(1)$ numbers.

We flesh out the details of our main argument in the next section and then go on to show the consequences for the dS conjecture in Sec-III before concluding in Sec-IV. From now on, we shall set the ($d$-dimensional) Planck mass $\Mpl$ to one for the rest of the paper. 

\section{TCC and SDC}
Let us assume a single tower of light states with equal spacing, say, for Kaluza-Klein (KK) modes. In this case the mass of light states appearing in the infinite tower goes as (some positive integer multiple of) 
\begin{eqnarray}\label{Dist}
m\(\varphi\) \sim e^{-\alpha\, |\Delta\varphi|}\,,
\end{eqnarray}
where $\alpha > 0$ is some $\mathcal{O}(1)$ number in general when $|\Delta\varphi| \gg 1$, according to the SDC \cite{Swampland1, SWC2, SWC3}. The refined version of the SDC states that this relation holds not only for moduli space but also for field space in an EFT with a potential $V \neq 0$. Our main takeaway from the SDC is that the number of light states in the tower (even for something as elementary as KK or winding modes) goes as 
\begin{eqnarray}\label{Nstar}
N_*\(\varphi\) \sim  e^{\alpha\, |\Delta\varphi|}\,,
\end{eqnarray}
where, once again, we assume a single massless tower. 

We now introduce the second ingredient for our derivation: When there is a large species of weakly-interacting particles, then the UV cut-off of the effective gravitational theory is given by \cite{SPB1}
\begin{eqnarray}\label{UV}
\Lambda_d \lesssim \frac{1}{N_*^{1/\(d-2\)}}\,,
\end{eqnarray}
where, from now on, we generalize to $d$-dimensions. This well-known result is the species scale conjecture which can be applied to the light states appearing in the SDC tower, as has been done in \cite{SDCTest1, dSfromSDC}. The number of light species below the true UV cut-off $\Lambda_d$ can be written as 
\begin{eqnarray}\label{N}
N_*\(\varphi\) \sim  \frac{\Lambda_d}{m_{KK}} = \Lambda_d\, e^{\alpha\, |\Delta\varphi|}\,.
\end{eqnarray}

From \eqref{N}, and using \eqref{UV},  we get as expression for the cut-off of the weakly-coupled gravitational regime as
\begin{eqnarray}\label{UV1}
\Lambda_d \lesssim e^{-\alpha\, |\Delta\varphi|/\(d-1\)} \,. 
\end{eqnarray}
For any EFT to be under perturbative control, we require that the curvature scale set by the Hubble parameter, corresponding to some energy density, be below this UV cut-off scale, \textit{i.e.}
\begin{eqnarray}\label{TCC}
\H < \Lambda_d  \;\;\Rightarrow \H < e^{-\alpha\, |\Delta\varphi|/\(d-1\)}\,.
\end{eqnarray}
This is similar to the bound as one finds for the TCC, when specialized to scalar fields \cite{TCC1}, for some $\mathcal{O}(1)$ number $\alpha$. At this point, it is pertinent to emphasize that the above upper bound is for the value of the Hubble parameter, $\H$, after the field has traversed a large geodesic distance on field space in order for the SDC to be valid.  From hereon, one can recover the TCC \eqref{TCCorg}, as follows.

We begin by rewriting \eqref{TCC} as
\begin{eqnarray}\label{TCC1}
\frac{\alpha\, |\Delta \varphi|}{\(d-1\)} < -\ln\(\H\)\,.
\end{eqnarray} 
From the $d$-dimensional Friedmann equation
\begin{eqnarray}\label{Friedmann}
\frac{\(d-1\) \(d-2\)}{2} H^2 = \frac{1}{2} \dot{\varphi}^2 + V\,,
\end{eqnarray}
it is easy to show that, for $V<0$, 
\begin{eqnarray}\label{TCC2}
\int_{\varphi_i}^{\varphi_f} \d \varphi \(\frac{H}{\dot{\varphi}}\) < \frac{|\Delta\varphi|}{\sqrt{\(d-1\) \(d-2\)}} < -\ln\(\H\)\,,
\end{eqnarray}
where we use \eqref{TCC1} to get the last inequality, which puts a lower bound on the exponent appearing from the SDC given by
\begin{eqnarray}
\alpha \geq \sqrt{\frac{d-1}{d-2}}\,.
\end{eqnarray}
But the left hand side of \eqref{TCC2} can be rewritten as the following to arrive at
\begin{eqnarray}\label{TCCfinal}
\int_{t_i}^{t_f}H \d t <  -\ln\(\H\)\,,
\end{eqnarray}
which is exactly the same as \eqref{TCCorg} when $\Mpl =1$.

Naturally, the more interesting case is for $V>0$ since this is the one required for (quasi-)dS backgrounds. In this case, we can use the equation for the rate of change of the Hubble parameter
\begin{eqnarray}\label{Hdot}
\dot{H}= - \(\frac{\dot{\varphi}^2}{d-2}\)\,
\end{eqnarray}
to rewrite the number of $e$-folds as
	\begin{eqnarray}
	\int_{\varphi_i}^{\varphi_f} \d \varphi \(\frac{H}{\dot{\varphi}}\) =  \frac{1}{\sqrt{\(d-2\)}}\int_{\varphi_i}^{\varphi_f} \d \varphi\, \frac{1}{\sqrt{\epsilon_H}} \simeq \frac{2}{\(d-2\)} \int_{\varphi_i}^{\varphi_f} \d \varphi\, \frac{V}{|V'|} \,,
	\end{eqnarray}
where $\epsilon_H := -\dot{H}/H^2$ and we have used the slow-roll approximation in the last equality. Next, using the lower bound on the quantity $|V'|/V$, as derived later in \eqref{Vprime_bound}, we get 
\begin{eqnarray}
\int_{\varphi_i}^{\varphi_f} \d \varphi \(\frac{H}{\dot{\varphi}}\)   < \(\frac{d-1}{d-2}\)\,\frac{1}{\alpha}\; |\Delta\varphi| \,.
\end{eqnarray}
Once again, using \eqref{TCC1} and rewriting the number of $e$-folds in a more familiar form, we see that 
\begin{eqnarray}\label{TCC_general}
\int_{t_i}^{t_f}H \d t < - \frac{\(d-1\)^2}{\(d-2\) \alpha^2} \, \ln\(\H\) \,.
\end{eqnarray}
This is equivalent to the TCC if we impose a stronger lower bound on the SDC exponent
\begin{eqnarray}\label{alpha}
\alpha \geq \frac{d-1}{\sqrt{d-2}}\,.
\end{eqnarray}

At this point, let us reiterate our main assumptions in deriving the TCC from the SDC and the species bound. Most crucially, we have to assume $|\Delta\varphi| \rightarrow \infty$ for using the lower bound \eqref{Vprime_bound}. However, it is natural to assume this in our case as this is the regime in which the SDC requires the emergence of the light tower of states. But it is important to emphasize that one cannot typically assume the upper bound for the number of $e$-folds used above in a regime where $|\Delta\varphi| \ll 1$. As has been emphasized right from the beginning, our derivation of the TCC is only valid in parametrically large distances on field space. We also emphasize that we are not secretly assuming the dS conjecture for our derivation to go through, which would indeed have been an additional requirement. Rather, the upper bound on $|V'|/V$ shall be derived in the next section, \textit{without invoking the TCC}, which is possible since one can arrive at the dS conjecture from the SDC on asymptotic regions of field space \cite{SWC1,dSfromSDC}. 

Our second assumption comes in the form of a derived bound on the $\mathcal{O}(1)$ constant in the exponent multiplying $|\Delta\varphi|$. We find that in order for the TCC to be valid, one must have the relation \eqref{alpha}. This is very similar to the lower bound found in \cite{TCC1} by comparing the distance conjecture and the TCC, up to $\mathcal{O}(1)$ numerical factors\footnote{Although small, this numerical factor can nevertheless play a crucial role in cosmological model-building, for instance, by becoming the deciding factor in ruling out quintessence models, compatible with the swampland, from observations \cite{QuintSwamp, QuintSwamp2, DESwamp}.}. As already pointed out in \cite{TCC1}, this opens up an opportunity to test this lower bound on $\alpha$ from explicit string theory constructions. Let us end this subsection with an interesting alternate possibility. Indeed, one can choose that $\alpha = \sqrt{\(d-1\)/\(d-2\)}$ such that \eqref{TCC1} acquires the same form as it did in \cite{TCC1}. The advantage of this would be that the lower bound on $|V'|/V > 2/\sqrt{\(d-1\)\(d-2\)}$  \eqref{Vprime_bound} remains the same as the one from \cite{TCC1}, as we shall show in the next section. This value of $\alpha$ would, of course, be unviable with our derived bound \eqref{alpha} above. However, one can interpret this as an alternate (weaker) version of the TCC, given by
\begin{eqnarray}\label{AlternateTCC}
\int_{t_i}^{t_f} H\d t < - \(d-1\)\, \ln\(\H\)\,,
\end{eqnarray}
which still puts a (weaker) upper bound the duration of the quasi-dS phase. Although this relation is quite similar to the TCC from this point of view, it is rather different if one takes the view that the fundamental property of the TCC is to stop trans-Planckian modes from crossing the Hubble radius. We just mention this to highlight the generality of the argument that the SDC, and the species bound, would always put an upper bound on the life-time of an accelerating phase, albeit the duration would depend on $\alpha \sim \mathcal{O}(1)$ \eqref{TCC_general}.

\section{TCC, dS conjecture and SDC}

Independently of the TCC, the dS conjecture also follows from our assumptions of starting from the SDC and using the species bound \cite{dSfromSDC}. Let us quickly repeat the main arguments of that derivation. Let us recall that $\H < \Lambda_d$. Moreover, since the potential must always satisfy $V < \frac{\(d-1\)\(d-2\)}{2} H^2$ from the Friedmann equation \eqref{Friedmann}, one gets an upper bound 
\begin{eqnarray}\label{Vbound}
V_f < A\, e^{-\frac{2 \alpha |\Delta\varphi|}{\(d-1\)}}\,,
\end{eqnarray}
which is a generalization of the Dine-Seiberg argument \cite{DS}. Here $A = \(d-1\) \(d-2\)/2$ is a constant. If one now chooses to take $|\Delta\varphi| \rightarrow \infty$, then we get the same conclusion as in \cite{dSfromSDC}, \textit{i.e.}, there cannot be any meta-stable dS vacua. This is how one arrives at the dS conjecture starting from the SDC.

With the benefit of hindsight (and of the arguments given in \cite{TCC1}), it can be seen that \eqref{TCC} is much more powerful and one can get a bound for the smallest initial value of $\varphi$, which then cannot start from arbitrary negative values to end up at $\varphi_f$. This is precisely why one finds that there are logarithmic corrections to the dS conjecture starting from the TCC, for short-range field excursions, as has nicely been explained in \cite{TCC1} (see also \cite{Geng:2019phi}). This helps in resolving a puzzle which might arise from our derivation -- how can the TCC and the dS conjecture both follow from our assumptions since the former allows for metastable dS vacua while the latter does not? It is simply because our argument is only valid for parametrically large distances on field space, in which regime, the TCC  agrees with the dS conjecture. 

However, if we were then to \textit{assume} that the TCC is valid everywhere on field space, then one gets the conclusion that a metastable dS vacua is allowed. Moreover, this also shows that under such an assumption, it would have been possible to derive the $log$-dependent correction factors one finds while deriving the dS conjecture from the TCC. In other words, it would have been possible to derive meta-stable dS vacua, along with a bound on their lifetimes as was shown in \cite{TCC1} (using \eqref{TCCfinal}), just by starting from the SDC and assuming the species bound, had one extended the findings to all of field space and not just restricted oneself to the asymptotic regions\footnote{It had been pointed out that there were some loopholes in the dS conjecture while deriving it in this way using the species bound and the SDC \cite{dSfromSDC}. Our argument offers a possible resolution to these loopholes in the sense that they perhaps only loosens the conjecture so as to allow metastable vacua with a given bound on their lifetimes.}. This can be made explicit as follows \cite{TCC1}: Starting from \eqref{Vbound}, one can write
\begin{eqnarray}\label{VprimeV}
\[\frac{-V'}{V}\]_{\varphi_i}^{\varphi_f} > \frac{1}{|\Delta\varphi|}\ln\( \frac{V_i}{A}\) + \frac{2 \alpha}{\(d-1\)}\,,
\end{eqnarray}
where the left hand side denotes the average of $\(V'/V\)$ over the field range $|\Delta\varphi|$.  As is obvious, the logarithmic terms disappear only for infinite field excursions to give us
\begin{eqnarray}\label{Vprime_bound}
\[\frac{-V'}{V}\]_\infty > \frac{2 \alpha}{\(d-1\)}\,.
\end{eqnarray}
Since we use this above bound for deriving the TCC, it is important to reiterate that we arrive at this version of the dS conjecture from the SDC, \textit{without assuming the TCC} at any point. As a final remark, it is easy to see that this matches with the bound in \cite{TCC1} when $\alpha = \sqrt{\(d-1\)/\(d-2\)}$, as mentioned in the previous section. 

Our derivation of the dS conjecture from the SDC plugs in another loophole in the original argument for the dS conjecture from entropy considerations on accelerating backgrounds \cite{SWC1}. In that case, one gets that the entropy of the light towers emerging due to the SDC, using the Bousso bound \cite{CEB1,CEB2}, scales as
\begin{eqnarray}\label{entropy}
S \propto e^{\alpha \varphi}\,.
\end{eqnarray}
Inverting the relation between the Gibbons-Hawking entropy for dS space and the potential, one finds that
\begin{eqnarray}
V \propto e^{-\alpha \varphi}\,,
\end{eqnarray}
where we ignore additional $\varphi$ dependence such as those arising from the density of light towers. However, as was pointed out in \cite{loophole1} (also see \cite{loophole2}), large flux compactifications can admit several balanced terms in the scalar potential even on parametrically large distances on field space, therefore also admitting a minima once such fluxes are turned on. An explicit counter-example given in \cite{loophole1} shows that for a potential of the form
\begin{eqnarray}\label{loophole}
V \sim A e^{-\alpha \varphi} + B e^{-\(\alpha +\beta\) \varphi} + C e^{-\(\alpha +\gamma\) \varphi}\,,
\end{eqnarray}
does admit a dS minima for $A \sim \mathcal{O}(1),\, B\sim - \mathcal{O}\(e^{\(\beta\varphi\)}\),\,  C\sim \mathcal{O}\(e^{\(\gamma\varphi\)}\)$, in the limit $\varphi \rightarrow \infty$. In other words, with flux-dependent coefficients of the right magnitude, one can get a dS solution thus stabilizing the Dine-Seiberg runaway. On the other hand, such a potential \eqref{loophole} still gives the same form for the entropy as expressed in \eqref{entropy}, near the minima, thereby reproducing the same expression as expected from the SDC. This, therefore, shows that by entropy arguments alone one \textit{cannot} derive of the dS conjecture starting from the SDC\footnote{In \cite{SWC1}, the entropy was further assumed to be strictly monotonically increasing, thereby avoiding this loophole.}. In our work, the upper bound for the potential \eqref{Vbound} comes from the SDC and the species bound argument alone without resorting to properties of quasi-dS backgrounds. Moreover, the lower bound on $\left[|V'|/V\right]_\infty$ is on the averaged value of $|V'|/V$ and not a local one. Therefore, even if one is to somehow assume the presence of some additional terms such that this upper bound can be ``stabilized'' by some flux compactifications (as in $B$ and $C$ terms of \eqref{loophole}), the resulting bound on $\left[|V'|/V\right]_\infty$, as given in \eqref{VprimeV}, is unaffected. This would lead to the presence of some additional terms in \eqref{VprimeV}; nevertheless in the limit $|\Delta\varphi|\rightarrow \infty$, one recovers \eqref{Vprime_bound}. 

\section{Conclusion}

There has been a lot of recent work aiming to see if inflation can be made compatible with the TCC. If correct, the TCC would imply a tectonic shift in our understanding of early-universe cosmology since it seems to highly disfavour models of inflation. Given its radical consequences, it is natural to ask if the TCC can be obtained from some other well-tested aspect of quantum gravity. 

In this work, we show how by only assuming the SDC, and a scaling of the UV cut-off for a large species of particles, one can arrive at the TCC for parametrically large distances in field space. Both the TCC, and its predecessor the `trans-Planckian' problem in inflation, were only motivated by EFT arguments without providing any quantum gravity support for it. In this work, we show how the TCC is \textit{not} a new swampland conjecture but rather a natural consequence of the well-established SDC. This implies that a consistent UV theory must secretly know that trans-Planckian modes should never exit the Hubble horizon. As a result, the nice thing about our logic is that we do not need to assume anything about quantum fluctuations crossing the Hubble horizon for our derivation. The price we pay is that our result is only valid in the asymptotic limit of large $|\Delta\varphi|$ and we need to conjecture that the TCC is valid more generally even for small field excursions. Contrary to previous approaches of assuming the TCC and examining its consequences, our result puts the TCC on a much firmer footing and makes it imperative to take its implications for inflationary cosmology \cite{TCC2}, among other things, more seriously. Interestingly, this establishes the TCC as a general expectation from consistent UV approaches, which has recently been further strengthened by deriving it using entropy arguments \cite{Sun:2019obt} or the Weak Gravity Conjecture \cite{Cai:2019dzj}\footnote{These papers arrived after the first version of this work appeared on arXiv.}.

\section*{Acknowledgments}
I thank Robert Brandenberger for many helpful discussions and for comments on an earlier version of this draft. I also thank Daniel Junghans, Gansukh Tumurtushaa and Shao-Jiang Wang  for interesting correspondence which helped improve the draft. This research is supported in part by funds from NSERC, from the Canada Research Chair program, by a McGill Space Institute fellowship and by a generous gift from John Greig.

\end{document}